\documentclass{ptapap}

\usepackage{array}
\usepackage{natbib}

\author{Marek Skarka}[ASU,KONK]
\author{Zden\v{e}k Prudil}[HEID]
\affil[ASU]{Astronomical Institue, Czech Academy of Sciences, Fri\v{c}ova 298, 251 65 Ond\v{r}ejov, Czech Republic}
\affil[KONK]{Konkoly Observatory, Research Centre for Astronomy and Earth Sciences, Hungarian Academy of Sciences, Konkoly Thege Miklós út 
15-17, H-1121 Budapest, Hungary}
\affil[HEID]{Astronomisches Rechen-Institut, Zentrum f\"{u}r Astronomie der 
Universität Heidelberg, Mönchhofstr. 12-14, D-691 20 Heidelberg, Germany}

\title{Photometric differences between modulated and  non-Blazhko RRab Lyrae stars in the Galactic bulge}
\headtitle{Differences between Blazhko and non-modulated RRab stars}

\begin{document}

\maketitle

\begin{abstract}

We present our results of searching for differences in light curves of modulated and non-modulated RRab stars in the Galactic bulge. We examined a sample of more than 8000 stars. The most important results are that Blazhko stars have shorter pulsation periods, less skewed mean light curves, lower mean amplitudes, larger rise-time, no difference in spatial distribution and metallicity.

\end{abstract}

\section{Introduction}\label{Sect:Introduction}

The amplitude and phase/period modulation of the light curve in RR Lyrae stars is known as the Blazhko (BL) effect \citep{blazhko1907}. The origin of the modulation has still not been satisfactorily explained \citep[see the overviews by][]{kovacs2016,smolec2016}. We aimed to search for possible differences in light-curve shape ($I$ filter) between BL and non-modulated stars in the fundamental mode RR Lyrae stars in the Galactic bulge \citep[observed by the OGLE survey,][]{soszynski2014}.

\section{Methods}\label{Sect:Methods} 
We employed classical Fourier decomposition techniques to identify the BL effect in the frequency spectra and describe the shape of the light curve via phase-independent Fourier coefficients \citep{simon1981}. The BL effect was searched among 8282 RRab stars with well-defined light curves brighter than 18\,mag \citep[for details see][]{prudil2017}.

\section{Results}\label{Sect:Results}
We identified modulation in 3341 RRab stars. From the comparison between light curves of modulated and non-BL stars, we found that BL stars have shorter pulsation periods on average, which is given by the decreasing number of BL stars with pulsation period above 0.6\,days. 

Modulated stars seem to be characteristic by small amplitudes and long rise times (RT), while among non-BL stars the situation is exactly the opposite. Consequently, Fourier amplitudes $R_{i1}$ are larger for stars with stable light curve, while phase parameters $\phi_{i1}$ are lower in comparison with modulated stars. These findings are given by the less skewed mean curves of modulated stars. The comparison between the parameters can be found in Table \ref{Tab:Parameters}. 

We also examined photometric physical parameters estimated on the basis of empirical formulae \citep[for example metallicity from][]{jurcsik1996}. Metallicity was found to be the same for BL and non-BL stars. Also the spatial distribution seems to be homogeneous and BL and non-BL stars are well mixed without preference to any direction or location in the Galactic bulge. More details can be found in \citet{prudil2017}.

From the $R_{31}$ vs. $\phi_{i1}$ diagrams it seems that two \citet{oosterhoff1939} groups can be present in the Galactic bulge \citep[see][this proceedings]{prudil2017b}.

\newcolumntype{M}{>{\footnotesize}c}
\begin{table}
\centering
\caption{Comparison of mean light-curve parameters of BL and non-BL stars from the Galactic bulge.}
{\footnotesize
\def\arraystretch{1.1}
\tabcolsep=3pt
\begin{tabular}{llllllllll}
\hline\hline
& $P$\,(days) & {\it Amp}\,(mag) & $R_{21}$ & $R_{31}$ & $\phi_{21}$\,(rad) & $\phi_{31}$\,(rad) & {\it RT} & [Fe/H]\,(dex) \\ \hline
non-BL & 0.564(1) & 0.577(3) & 0.491(1) & 0.319(1) & 4.450(4) & 2.832(7) & 0.169(3) & -0.962(5) \\
BL & 0.533(1) & 0.540(3) & 0.458(1) & 0.274(1) & 4.367(4) & 2.616(8) & 0.219(8) & -0.969(5) \\ \hline

\hline
\end{tabular}\label{Tab:Parameters}
}
\end{table}

\acknowledgements{MS acknowledges the financial support of the Czech Grant GA \v{C}R 17-01752J, the NKFIH-115709 grant of the Hungarian National Research, Development and Innovation office, and the support of the postdoctoral fellowship programme of the Hungarian Academy of Sciences at the Konkoly Observatory as host institution. ZP acknowledges the support of the Hector Fellow Academy.}


\end{document}